\documentclass[amsmath]{revtex4}
\usepackage{longtable}
\usepackage{graphicx}
\usepackage{times}
\usepackage{amssymb}
\newcommand{\beq}{\begin{equation}}
\newcommand{\eeq}{\end{equation}}
\newcommand{\bdm}{\begin{displaymath}}
\newcommand{\edm}{\end{displaymath}}

\begin{document}

\title{The finite mass beamsplitter in high power interferometers}

\author{Jan Harms}
\author{Roman Schnabel}
\author{Karsten Danzmann}

\affiliation{Institut f\"ur Atom- und Molek\"ulphysik, Universit\"at Hannover and Max-Planck-Institut f\"ur Gravitationsphysik (Albert-Einstein-Institut), Callinstr.~38, 30167 Hannover, Germany}

\date{\today}

\begin{abstract}The beamplitter in high-power interferometers is subject to significant radiation-pressure fluctuations. As a consequence, the phase relations which appear in the beamsplitter coupling equations oscillate and phase modulation fields are generated which add to the reflected fields. In this paper, the transfer function of the various input fields impinging on the beamsplitter from all four ports onto the output field is presented including radiation-pressure effects. We apply the general solution of the coupling equations to evaluate the input-output relations of the dual-recycled laser-interferometer topology of the gravitational-wave detector GEO\,600 and the power-recycling, signal-extraction topology of advanced LIGO. We show that the input-output relation exhibits a bright-port dark-port coupling. This mechanism is responsible for bright-port contributions to the noise density of the output field and technical laser noise is expected to decrease the interferometer's sensitivity at low frequencies. It is shown quantitatively that the issue of technical laser noise is unimportant in this context if the interferometer contains arm cavities.
\end{abstract}

\maketitle 

\section{Introduction}
Earth-bound laser-interferometers seeking for gravitational waves \cite{geo02,LIGO,TAMA,VIRGO} use high-power light fields in order to minimise the quantum-noise in the detection band. The main contribution to the quantum noise comes from the output port itself \cite{Cav80}. The output port vacuum field is reflected by the interferometer back towards the photodetector. Its noise spectral density was studied in great detail \cite{KLMTV01,BCh01}. One can manipulate the dark port field, thereby increasing the sensitivity of the gravitational-wave detection in the frequency band of interest. It was proposed by Caves to squeeze the vacuum field \cite{Cav81}. Combined with an appropriate filtering scheme, the increase in sensitivity is limited by the squeezing factor. This was shown for different interferometer topologies \cite{KLMTV01,JH03,Sch04}. In all these investigations radiation-pressure noise at the beamsplitter was not included. 

In this paper we analyze the effect of radiation-pressure fluctuations acting on the beamsplitter. We show that it gives rise to a coupling of the bright input port to the dark output port. Consequently this paper focusses on the field which enters the interferometer at the bright port, i.~e.~the input light which comes from the laser. We quantitatively investigate the contributions of quantum noise and technical laser noise of the bright input port to the interferometer's noise spectral density of the output field. We show that the coupling strongly depends on the interferometers topology and that technical laser noise might limit the detectors sensitivity at low frequencies.

The paper is organized as follows. In section II, we discuss some general properties of beamsplitters and introduce the coupling equations of the fields. The solution of the coupling equations is presented in section III for a specific configuration, the power-recycled interferometer operating at dark fringe. In section IV, we present the noise spectral density for the current set-up of the dual-recycled GEO\,600 interferometer and also for its envisioned design parameters. We conclude that in the design configuration of GEO\,600 the dark port noise spectral density might be dominated by technical noise from the bright port at low frequencies. In section V, the same calculations are performed for the advanced LIGO configuration. A quantitative comparison shows that the relative contribution of the bright port noise to the output spectral density for advanced LIGO is smaller by five orders of magnitude than for GEO\,600.

\section{The Coupling Equations}
Quantum fields are usually described by means of their annihilation and creation operators. The two-photon formalism developed by Caves and Schumaker \cite{CSc85} turns out to be a more suitable formalism for measurements with heterodyne or homodyne detectors. These two classes of detectors measure the quadrature fields of the light whose amplitudes annihilate quanta of modulations. Correlations between the two sidebands built up by two-photon processes find a natural representation in that formalism and the spectral densities of the two quadratures' quantum noise is given by orthogonal sections through the so called noise ellipse. Modern publications discussing high-power interferometry show that one can derive simple and easy-to-interpret expressions for the quadrature transfer functions of various configurations \cite{KLMTV01,BCh01,BCh02,Ch03,JH03,PC02}. 
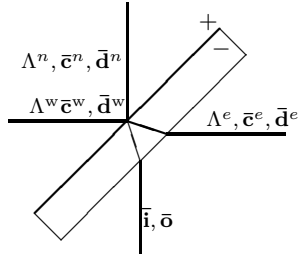
\begin{figure}[ht]
\begin{picture}(100,100)			
\thicklines
\put(10,20){\line(1,1){70}}
\thicklines
\put(0,55){\line(1,0){45}}
\put(45,55){\line(3,-1){15}}
\put(60,50){\line(1,0){45}}
\put(45,55){\line(0,1){45}}
\thinlines
\put(45,55){\line(1,-3){5}}
\put(50,40){\line(0,-1){35}}
\put(72,90){$+$}
\put(77,80){$-$}
\footnotesize
\put(8,58){$\Lambda^{\rm w}{\bf\bar c}^{\rm w},{\bf\bar d}^{\rm w}$}
\put(75,53){$\Lambda^e,{\bf\bar{c}}^e,{\bf\bar{d}}^e$}
\put(5,75){$\Lambda^n,{\bf\bar{c}}^n,{\bf\bar{d}}^n$}
\put(51,15){$\bf\bar{i},\bf\bar{o}$}
\put(20,10){\line(1,1){70}}
\put(20,10) {\line(-1,1){10}}
\put(80,90){\line(1,-1){10}}
\end{picture}
\caption{The noise density of the output field ${\bf\bar o}$ determines the noise of the gravitational-wave detection. All ${\bf\bar c}^i$ and the input field ${\bf\bar i}$ are propagating towards the beamsplitter. The fields ${\bf\bar d}^i$, ${\bf\bar o}$ propagate away from it. The asymmetric beamsplitter reflects with a minus sign on the side where it is indicated in the picture. The $\Lambda^i$ denote the classical amplitudes of the carrier in each direction. The south port does not contain any carrier field at dark fringe.}
\label{split}
\end{figure}
Therefore, we present all equations in the two-photon formalism benefitting from algebraic properties of the quadrature fields concerning radiation-pressure effects. The two quadrature amplitudes $\hat{a}_1$, $\hat{a}_2$ merge into one single object which we call the quadrature vector:
\beq
{\bf\bar a} = \begin{pmatrix} \hat{a}_1 \\ \hat{a}_2 \end{pmatrix}
\eeq
Many important physical transformations acting on ${\bf\bar a}$ can be interpreted geometrically as rotations and scalings of vectors in the space spanned by the two quadrature fields (i.e. the quadrature space). A more detailed treatment is given in \cite{HarmsDipl}. Our notational conventions are introduced in Fig.~\ref{split}. The classical carrier amplitudes are treated seperately from the modulation amplitudes. Therefore, we assume that the expectation values of both components of all quadrature vectors are much smaller than their carrier amplitudes $\Lambda^i$. In the literature, one finds different conventions for the phase relations of the various fields which couple at the beamsplitter. However, for a lossless beamsplitter one can derive Stokes-like reciprocity relations involving the reflection and transmission of light which require that the amplitudes couple at the beamsplitter according to
\begin{eqnarray}
{\bf\bar o} & = & \tau {\bf\bar c}^n-\rho P_x{\bf\bar c}^e \nonumber \\
{\bf\bar d}^e & = & \tau {\bf\bar c}^w-\rho P_x{\bf\bar i} \nonumber \\
{\bf\bar d}^n & = & \tau {\bf\bar i}+\rho P_{-x}{\bf\bar c}^w \nonumber \\
{\bf\bar d}^w & = & \tau {\bf\bar c}^e+\rho P_{-x}{\bf\bar c}^n
\label{coupling}
\end{eqnarray}
An explicit expression for $P_x$ is developed in the next section, when the power-recycled Michelson interferometer is discussed. At this point, we state some of its general features. The operator $P_x$ counts for the change of phase relations due to displacements $\hat{x}$ of the beamsplitter. Therefore, $P_x$ can be thought of as a propagator of the field along $\hat{x}$ corresponding to a rotation of the field's quadrature vector in quadrature space. Since the coupling equations relate modulation amplitudes, $\hat{x}$ denotes the amplitude for displacements of the beamsplitter at some frequency $\Omega$ which is the modulation frequency of the field. Gravitational waves do not affect the motion of the beamsplitter in its own proper reference system and consequently $\hat{x}$ is independent of the gravitational-wave amplitude $h$. In that case, the equation of motion for $\hat{x}$ is completely determined by the radiation-pressure fluctuations of the light, i.e. by the fluctuations of the amplitude quadratures $\hat{a}_1$ of all the fields. We derive the equation of motion in terms of momentum conservation. The beamsplitter has to compensate for the momentum flow of the ingoing and outgoing fields. Therefore, we make the following linearized ansatz in terms of the modulation amplitudes
\beq
\hat{x}\propto \Lambda^w(\hat{c}_1^w+\hat{d}_1^w)-\Lambda^e(\hat{c}_1^e+\hat{d}_1^e)+\Lambda^n(\hat{c}_1^n+\hat{d}_1^n).
\label{motion}
\eeq
The minus sign in front of the second bracket means that the momentum assigned to the east fields is carried in the opposite direction with respect to the momentum carried by the west fields, whereas the plus sign in front of the last bracket means that a motion of the beamsplitter downwards is equivalent to a motion towards the east concerning phase shifts of the reflected light. The square root of the spectral density of position fluctuations $\sqrt{S(\hat{x})}$ is supposed to be much smaller than the wavelength $\lambda_0$ of the carrier light. If that condition were not fulfilled, then the backaction of our measurement device on the test masses would be much larger than typical displacements induced by a gravitational wave ($x\approx 10^{-10}\cdot\lambda_0$, depending on the amplitude of the gravitational wave). If the field which the propagator $P_x$ acts on is not accompanied by a high-power carrier amplitude $\Lambda$, then the propagation becomes the unity matrix. In other words, the position fluctuations of the beamsplitter do not generate any sidebands, because there is no carrier on which sidebands with significant amplitude could be modulated. Then we obtain the following expression for the propagation along small displacements if the interferometer operates at dark fringe
\begin{eqnarray}
P_x{\bf\bar i} & = & {\bf\bar i} \nonumber \\
P_x{\bf\bar c}^j & = & {\bf\bar c}^j-\Lambda^j\cdot{\bf\bar \kappa}({\bf\bar c}^i,{\bf\bar d}^i)
\label{prop}
\end{eqnarray}
The vector ${\bf\bar\kappa}$ is a linear function of $\hat{x}$ and consequently, it depends on all the fields which enter into the equation of motion of the beamsplitter. It should be clear that we just need one variable to determine the position of the beamsplitter since, concerning the phase shift of the reflected fields, a motion of the beamsplitter downwards is completely analogous to a motion to the right.

We are going to add three more equations to our system of coupling relations Eq.~(\ref{coupling}). The idea is to assign roundtrip transfer functions $E,N,W$ and independent fields ${\bf\bar e},{\bf\bar n},{\bf\bar w}$ to three of the four ports. The new fields comprise a sum of all fields originating in the corresponding port , e.g. vacuum fields due to losses or classical signal fields due to a gravitational wave. One may understand this step as some sort of closure of the ports by means of mirrors which reflect the outgoing light back to the beamsplitter. 
\begin{eqnarray}
{\bf\bar c}^e & = & E{\bf\bar d}^e+ {\bf\bar e} \nonumber \\
{\bf\bar c}^n & = & N{\bf\bar d}^n+{\bf\bar n} \nonumber \\
{\bf\bar c}^{\rm w} & = & W{\bf\bar d}^{\rm w}+{\bf\bar w}
\label{ports}
\end{eqnarray}
The latter equations are the most general, linear equations which govern the roundtrip of the light. In the two-photon formalism, the transfer functions $E,N,W$ are transfer matrices acting on quadrature vectors. 

\section{THE INPUT-OUTPUT RELATION}
The input-output relations of an optical system comprise all contributions to the output field, i.~e.~the field which is detected by the photodiode. It is obtained by solving the coupling equations Eq.~(\ref{coupling}) and Eq.~(\ref{ports}):
\beq
{\bf\bar o}={\bf IO}({\bf\bar i},{\bf\bar w},{\bf\bar n},{\bf\bar e})
\eeq
\begin{figure}[ht]
\begin{picture}(100,100)			
\put(10,30){\line(1,0){45}}
\put(40,10){\line(0,1){35}}
\put(35,45){\framebox(10,35){$A$}}
\put(55,25){\framebox(35,10){$A$}}
\thicklines
\put(35,25){\line(1,1){10}}
\put(10,25){\line(0,1){10}}
\footnotesize
\put(5,38){$\rho_{\rm pr}$}
\put(20,32){$W$}
\end{picture}
\caption{Power-recycled interferometer. Both arms are described by the same transfer matrix $A$. The west port contains a power-recycling mirror with amplitude reflectivity $\rho_{\rm pr}$. It forms the power-recycling cavity with the endmirrors of the Michelson interferometer. The mirror's distance to the beamsplitter is set to be an integer multiple of the carrier wavelength. The same holds for the pathlength of the light inside the two interferometer arms.}
\label{inter}
\end{figure}
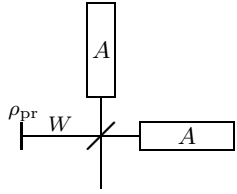
We present the solution of the coupling relations for a power-recycled interferometer with a 50/50 beamsplitter operating at dark fringe as shown in Fig.~(\ref{inter}):
\begin{eqnarray}
& \rho=\tau=\frac{1}{\sqrt{2}} & \quad \mbox{50/50 beamsplitter} \nonumber \\
& \Lambda:=\Lambda^w=\sqrt{2}\Lambda^e=\sqrt{2}\Lambda^n & \quad \nonumber \\
& A:=E=N & \quad \mbox{dark fringe condition} \nonumber \\
& W:=\rho_{\rm pr}(1+{\rm RPE}) & \quad \mbox{power recycling}
\end{eqnarray}
The power-recycling condition means that the transfer function $W$ is a multiple of the identity map except for radiation-pressure effects. Without loss of generality, our condition requires that the distance of the beamsplitter to the power-recycling mirror is a multiple of the carrier wavelength, which also implies that the pathlength of the light inside the Michelson arms is a multiple of the carrier wavelength. In fact, the proper power-recycling condition is weaker than the one imposed here for simplicity. The proper condition merely requires that the combined pathlength through the Michelson arm and the power-recycling cavity is a multiple of the carrier wavelength. The radiation-pressure induced noise sidebands generated at the beamsplitter are derived from the matrix for small propagations and from the equation of motion of the beamsplitter. Small propagations $P_x$ lead to the following transformation of the quadrature vectors \cite{HarmsDipl}:
\beq
\begin{pmatrix} 1 & -\frac{\omega_0}{c}\hat{x} \\ \frac{\omega_0}{c}\hat{x} & 1\end{pmatrix}{\bf\bar c}^j \approx{\bf\bar c}^j+\frac{\omega_0 \Lambda^j}{c}\begin{pmatrix} 0 \\ \hat{x}({\bf\bar c}^i,{\bf\bar d}^i) \end{pmatrix}
\label{side}
\eeq
Implicitly, we made use of the fact that ${\bf\bar c}^j$ is a modulation amplitude of a carrier field whose amplitude $\Lambda^j$ points in the direction of the amplitude quadrature of ${\bf\bar c}^j$. The second term on the right-hand side corresponds to the noise sidebands which are excited by fluctuations of the phase $\omega_0\hat{x}/c$. In the two-photon formalism, phase fluctuations yield fluctuations of the phase quadrature whose noise amplitude is the phase shift multiplied by the amplitude of the carrier field. We assumed that the carrier frequency $\omega_0$ is much higher than the modulation frequency which, henceforth, is denoted by $\Omega$. The equation of motion for $\hat{x}$ is governed by Newton's law
\beq
\hat{x}=-\frac{\delta P({\bf\bar c}^i,{\bf\bar d}^i)}{mc\Omega^2}
\eeq
Here, Newton's equation is written in the domain of modulation frequencies and $m$ is the mass of the beamsplitter. The fluctuating part $\delta P$ of the radiation pressure is proportional to the right-hand side of Eq.~(\ref{motion}). The constant of proportionality can be determined by comparing our expression for $\delta P$ with expressions evaluated for simpler geometrical situations (e.~g.~see \cite{KLMTV01}):
\beq
\delta P=\hbar\omega_0\cdot\left[\Lambda^w(\hat{c}_1^w+\hat{d}_1^w)-\Lambda^e(\hat{c}_1^e+\hat{d}_1^e)+\Lambda^n(\hat{c}_1^n+\hat{d}_1^n)\right]
\eeq
Bringing everything together, we cast Eq.~(\ref{side}) into the form
\begin{widetext}
\beq
P_x{\bf\bar c}^j \approx {\bf\bar c}^j -\frac{\hbar\omega_0^2\Lambda\Lambda^j}{mc^2\Omega^2}\begin{pmatrix} 0 & 0 \\ 1 & 0 \end{pmatrix}\cdot\left[({\bf\bar c}^w+{\bf\bar d}^w)-\frac{1}{\sqrt{2}}({\bf\bar c}^e+{\bf\bar d}^e)+\frac{1}{\sqrt{2}}({\bf\bar c}^n+{\bf\bar d}^n)\right]
\label{Pnice}
\eeq
\end{widetext}
where $\Lambda$ denotes the amplitude of the light in the west port. Before we write down the input-ouput relation, we introduce the abbreviation
\beq
K=\begin{pmatrix} 0 & 0 \\ -K_B & 0 \end{pmatrix},\quad K_B=\frac{\hbar\omega_0^2\Lambda^2}{mc^2\Omega^2}
\label{beamC}
\eeq
The coupling constant $K_B$ is proportional to the power of the light at the beamsplitter by virtue of $P=\hbar\omega_0\Lambda^2$. Inserting Eq.~(\ref{Pnice}) into the coupling equations and subsequently solving the system of linear equations for the output field, one obtains
\begin{eqnarray}
{\bf\bar o} & = & \frac{1}{2(1-A\rho_{\rm pr})-(2+(1-A)\rho_{\rm pr})(1+A)K}\Big[ \nonumber \\
& & \quad \left(2A(1-A\rho_{\rm pr})+(1-(1+2\rho_{\rm pr})A)(1+A)K\right){\bf\bar i} \nonumber \\
& & \quad +\sqrt{2}(1-A\rho_{\rm pr}){\bf\bar n} \nonumber \\
& & \quad -\sqrt{2}((1-A\rho_{\rm pr})-(1+\rho_{\rm pr})(1+A)K){\bf\bar e} \nonumber \\
& & \quad +(1+A)^2K{\bf\bar w}\Big]
\label{iorel}
\end{eqnarray}
If the radiation-pressure fluctuations are negligible, then the matrix $K$ becomes zero and the input-output relations reduce to a well-known form. The most interesting aspect of this result is probably contained in the last term within the square brackets. It says that whenever there are radiation-pressure fluctuations acting on the beamsplitter, then fluctuations from the west port (also known as the bright port) couple to the output port. This contribution is proportional to the non-zero component of the matrix $K$. This might turn out to be a problem for all high-power interferometers, since the laser field suffers from high technical noise at low sideband frequencies, which couples into the field ${\bf\bar w}$. The technical noise at low frequencies can be several orders higher compared to pure vacuum fluctuations. The fact that the bright-port dark-port coupling is proportional to $K$ also explains why the input-output relations are independent of radiation-pressure fluctuations acting on the power-recycling mirror. Those fluctuations are described by a matrix $K^\prime$ which has the same form than $K$ and the transfer is governed by multiplying that matrix with $K$ and $K\cdot K^\prime$ is always zero. 

\section{THE NOISE SPECTRAL DENSITY OF THE GEO\,600 TOPOLOGY}
\begin{figure}[ht!]
\centerline{\includegraphics[width=8cm]{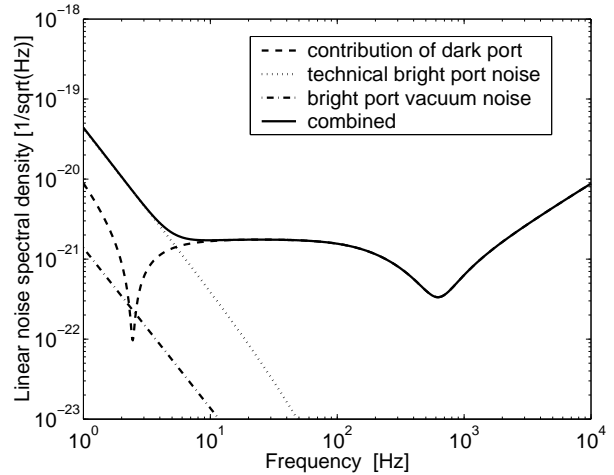}}
\caption{Single-sided noise spectral density of GEO\,600 with $P=300\,$W at the beamsplitter. The spectral density of the bright port vacuum field is lying below the dark port noise spectral density throughout the entire detection band. However, the technical noise from the bright port is dominating the spectral density up to $10\,$Hz where it is more than one order of magnitude higher than the vacuum noise density. The technical noise corresponds to an input laser field with power $P_{\rm in}=1.5\,$W .}
\label{dens300}
\end{figure}

\begin{figure}[ht!]
\centerline{\includegraphics[width=8cm]{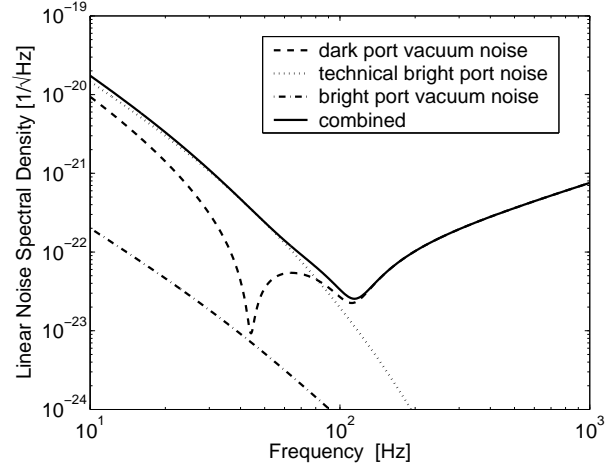}}
\caption{Single-sided spectral density of the dark-port field for the GEO\,600 topology with design power $P=10\,$kW at the beamsplitter assuming the same relative technical noise here than for the low power interferometer underlying Fig.~\ref{dens300}. However, the absolute technical noise of the input field is increased due to the higher input power of the light $P_{\rm in}=10\,$W.}
\label{dens10k}
\end{figure}

In this section, we calculate the input-output relations of the dual-recycled configuration of GEO\,600 and evaluate them in terms of the noise spectral density which is obtained under the following assumptions. The state of the input field $\bf\bar i$ at the south port is a coherent vacuum field and $\bf\bar w$ is the fraction of the laser field which transmits into the power-recycling cavity. Expressed in terms of single-sided spectral density matrices these properties assume the form
\beq
{\bf S}({\bf\bar i})={\bf 1}, \quad
{\bf S}({\bf\bar w})=\tau_{\rm pr}^2\cdot{\bf S}_{\rm tech}(\Omega)
\eeq
The matrix ${\bf S}_{\rm tech}$ is diagonal which means that our calculations do not account for correlations between the two quadratures built up inside the laser. The amount of technical noise which is brought into the interferometer by $\bf\bar w$ is estimated from measurements performed on the GEO\,600 laser. Optical losses occuring in real interferometers at the endmirrors or at the beamsplitter are not included in the sense that we do not mix the fields inside the interferometer with loss related vacuum fields. The value of the classical amplitude of the carrier light at different points of the interferometer is taken from real measurements. 
\begin{figure}[ht!]
\centerline{\includegraphics[angle=-90,width=8cm]{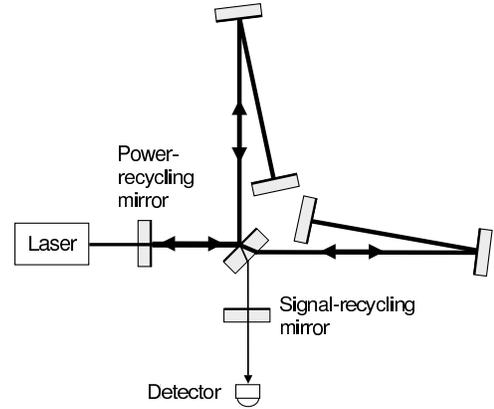}}
\caption{GEO\,600 is a dual-recycled Michelson interferometer with a power-recycling mirror in the bright port that enhances the light power within the Michelson arms and a signal-recycling mirror in the dark port that can be tuned to a specific signal frequency. Since the arms are folded once, the effective armlength is doubled to 1200~m. The distance between the beampslitter and the so called far mirrors of the Michelson arms is $600\,$m, whereas the so called near mirrors which form the end of the arms are placed very close to the beamsplitter.}
\label{GEO600}
\end{figure}
The equations of motion of all optical components are determined by the light pressure and the action of a gravitational wave. The latter one couples to the fields $\bf\bar n$ and $\bf\bar e$. No significant signal is found in the field $\bf\bar w$ since the distance of the power-recycling mirror to the origin of our reference frame (i.~e\;the beamsplitter) is small compared to the lengths of the two Michelson arms. A transfer matrix $A$ for the arms was first presented in \cite{KLMTV01} and was derived in \cite{HarmsDipl} for the GEO\,600 configuration applying the same formalism:
\beq
A=e^{2i\frac{\Omega L}{c}}\begin{pmatrix} 1 & 0 \\ -K_A & 1 \end{pmatrix}
\label{arm}
\eeq
The optomechanical coupling constant $K_A$ of the Michelson arms is defined similarly to the beamsplitter coupling constant $K_B$ in Eq.~(\ref{beamC}) with the amplitude $\Lambda$ substituted by the amplitude of the light inside the arms and the beamsplitter mass $m$ substituted by the reduced mass for the two endmirrors (each having mass $m_M$) which form the folded arms of GEO\,600 (see Fig.~\ref{GEO600}):
\beq
K_A=4\cdot\frac{\hbar\omega_0^2(\Lambda^2/2)}{(m_M/5)c^2\Omega^2}
\label{couK}
\eeq
By folding the arms, the effective armlength $L$ becomes twice the distance between the far mirror and the beamsplitter. A gravitational wave $h$ creates signal sidebands in both arms which possess equal amplitudes but different signs
\beq
{\bf\bar n}=-{\bf\bar e}=e^{i\frac{\Omega L}{c}}\frac{\sqrt{K_A}}{h_{\rm SQL}} {0 \choose h},\quad h^2_{\rm SQL}=\frac{4\hbar}{(m_M/5)\Omega^2L^2}
\label{signal}
\eeq
The quantity $h_{SQL}$ is the standard quantum limit of GEO\,600 with an infinite mass beamsplitter. The ''true'' quantum limit for GEO\,600 also depends on the dynamics of the beamsplitter. We refrain from redefining $h_{SQL}$ in that manner, since here we want to discuss the beamsplitter dynamics explicitely and we do not want to find the reduced mass motion of the system. The problem to calculate the phase and coupling constant of a folded arm transfer function is related to the calculation of the same quantities for a delay line. A nice treatment of delay lines in our formalism can be found in the appendix of \cite{Ch03}. For this particular set of matrices [Eqs.~(\ref{arm}),(\ref{signal})], the input-output relation Eq.~(\ref{iorel}) is given by
\beq
{\bf\bar o}=e^{2i\frac{\Omega L}{c}}\begin{pmatrix} 1 & 0 \\ -K_1 & 1 \end{pmatrix}{\bf\bar i}+e^{2i\frac{\Omega L}{c}}\begin{pmatrix} 0 & 0 \\ -K_2 & 0 \end{pmatrix}\,{\bf\bar b}+\sqrt{2}\,{\bf\bar n}
\label{GEOio}
\eeq
We substituted the field $\bf\bar w$ by the transmitted bright port input field ${\bf\bar w}=\tau_{\rm pr}{\bf\bar b}$. The two constants $K_1$ and $K_2$ depend on the arm and beamsplitter coupling constants
\begin{eqnarray}
K_1 & = & K_A+2\cos^2\textstyle{(\frac{\Omega L}{c})}\cdot K_B, \nonumber \\ 
K_2 & = & 2\cos^2\textstyle{(\frac{\Omega L}{c})}\displaystyle{\frac{\tau_{\rm pr}}{1-\rho_{\rm pr}e^{2i\frac{\Omega L}{c}}}}\cdot K_B
\label{couplings}
\end{eqnarray}
The coupling constant $K_2$ is a product of $2\cos^2(\frac{\Omega L}{c})\approx 2$ and the amplification factor for modulation fields inside the power-recycling cavity. We should emphasize that $K_2$ is independent of the arm coupling constant $K_A$ and thus independent of the arm topology (i.~e.\;whether it is a Michelson interferometer without arm cavities or with arm cavities). However, the modulus of the quantity $K_2$ decreases if the arm length $L$ is increased. From Eq.~(\ref{GEOio}) one derives the input-output relation of the signal-recycled interferometer in the usual manner. Propagating fields from the beamsplitter to the signal-recycling mirror is accomplished by a rotation matrix $D(\phi)$ acting in quadrature space which lacks the additional phase shift of the modulation fields since the wavelength $\lambda=(2\pi c)/\Omega$ of the sidebands within the detection band (i.~e.\;$10\,$Hz-$1000\,$Hz) is much longer than the length of the signal-recycling cavity \cite{BCh01}
\beq
D(\phi)=\begin{pmatrix} \cos(\phi) & -\sin(\phi) \\ \sin(\phi) & \cos(\phi) \end{pmatrix}
\eeq
The angle $\phi$ is the detuning parameter of the signal-recycling cavity which is formed by the signal-recycling mirror and the Michelson interferometer. In Eq.~(\ref{GEOio}), giving names $T_i$ and $T_b$ to the transfer matrices of the fields $\bf\bar i$ and $\bf\bar b$ respectively, the input-output relation for the signal-recycled interferometer reads
\begin{eqnarray}
{\bf\bar o}_{sr} & = & \frac{\bf 1}{{\bf 1}-\rho_{sr}\cdot D(\phi)T_iD(\phi)}\Big[\Big(D(\phi)T_iD(\phi)-\rho_{sr}\cdot{\bf 1}\Big){\bf\bar i}_{sr} \nonumber \\
& & \qquad +\tau_{sr}\,D(\phi)\cdot\Big(T_b\,{\bf\bar b}+\sqrt{2}\,{\bf\bar n}\Big)\Big]
\end{eqnarray}
The input-output relation determines the noise spectral density of the output field. The overall noise density is a sum of the two densities for the input field ${\bf\bar i}_{sr}$ and $\bf\bar b$. The latter one is the technical noise transferred from the bright port, the former one is the vacuum noise reflected at the dark port. It is convenient to normalize the spectral densities of the amplitude and phase quadratures of ${\bf\bar o}_{sr}$ such that the spectral density refers to the amplitude $h$ of the gravitational wave which is contain in $\bf\bar n$. The way how to do this normalization in matrix notation is shown in \cite{JH03}. The evaluation of the spectral density is based on the parameter values according to Table \ref{params}.  
\begin{table}[ht!]
\begin{tabular}{l|cr}
\hline\hline
& Symbol & Value $\quad$\\
\hline
Light power at BS & $P$ & $300\,\rm W\quad$\\
Transmissivity PRM $\quad$ & $\qquad\tau_{\rm pr}^2\qquad$ & $\quad 1.35\%\quad$ \\
Transmissivity SRM & $\tau_{\rm sr}^2$ & $2\%\quad$ \\
Beamsplitter mass & $m$ & $9.3\,\rm kg \quad$ \\
Mirror mass & $m_M$ & $5.6\,\rm kg \quad$ \\
Arm length & $L$ & $1200\,\rm m \quad$ \\
Frequency of laser & $\omega_0$ & $1.77\cdot 10^{15}\,{\rm rad/s}\quad$ \\
Detuning of SR cavity & $\phi$ & $0.015\,\rm rad\quad$ \\
\hline\hline
\end{tabular}
\caption{Parameters of the GEO\,600 configuration during the S3 run \cite{S3run}. The detuning $\phi$ of the signal-recycling cavity can be varied. The input light power at the power-recycling mirror was about $1.5$W.}
\label{params}
\end{table}
A detuning $\phi=0.015$ means that the sideband which lies $600\,$Hz above the carrier is resonantly amplified within the signal-recycling cavity. Adjusting the phase of the local oscillator in a homodyne detection scheme (corresponding to the electronic demodulation phase in heterodyne detection schemes), one can choose the direction in quadrature space along which the measurement is carried out. In that manner, the phase quadrature, the amplitude quadrature or some intermediate linear combination of these two can be measured. We refer to \cite{BChMa} for a deeper discussion of the quantum noise in heterodyne measurement schemes. Here, we restrict to measurements of the phase quadrature. The single-sided noise spectral density of the phase quadrature of the output field is shown in Fig.~\ref{dens300}.  
The bright port noise at low frequencies causes the optomechanical resonance to disappear from the noise spectral density. On the one hand, this effect is merely of theoretical interest as the currently measured noise density at low frequencies is dominated by seismic noise which couples to the optical fields through the mirror suspension. On the other hand, the result suggests that one has to investigate the role of bright port fluctuations for future interferometers. The beamsplitter coupling constant $K_B$ is proportional to the light power at the beamsplitter. Therefore, one might expect that the transferred bright port noise becomes even more significant for high power interferometers of the next generation. The corresponding noise spectral density for GEO\,600 with design power $P=10\,$kW and adjusted detuning $\phi=0.003$ and transmissivity $\tau_{\rm sr}^2=0.16\%$ is shown in Fig.~\ref{dens10k} assuming the same relative technical noise than before.
Since the power of the carrier light is higher than in the previous case, all coupling constants are increased and the low frequencies noise experiences a shift upwards. Furthermore, the absolute technical bright-port noise was scaled by a factor 10\,W/1.5\,W derived from the two respective input powers.

We conclude this section by suggesting a quantity which best characterizes the impact of the bright-port dark-port coupling on the output spectral density. That quantity should describe the balance of contributions coming from the bright port and the dark port to the output noise. We are looking for a characteristic function of the interferometer topology which is independent of the input power. We derive such a quantity from the input-output relation Eq.~(\ref{GEOio}) by  comparing the components of the transfer matrix of the bright port field with the components of the matrix for the dark port field. At low frequencies (i.~e.\;less than $10\,$Hz) it suffices to compare the values of the coupling constants $K_1$ and $K_2$ defined in Eq.~(\ref{couplings}). Their ratio $|K_1|/|K_2|$ tells us which field mainly determines the fluctuations in the output field $\bf\bar o$. If the ratio is bigger than one, then the dark port field $\bf\bar i$ dominates. If the ratio is less than one, then the bright port field $\bf\bar b$ dominates. We call this ratio the low frequency balance and denote it by $\sigma$. For GEO\,600 without signal-recycling mirror one obtains the following expression:
\beq
\sigma_{\rm conv} \approx\frac{\tau_{\rm pr}}{2}\left(\frac{K_A}{2K_B}+1\right)=\frac{\tau_{\rm pr}}{2}\cdot\frac{m+m_M/5}{m_M/5}\approx 0.5
\label{estCONV}
\eeq
This equation states that the bright-port fluctuations at low frequencies contribute twice as much as the dark-port fluctuations. Comparing the result with the noise spectral density of the vacuum fields in Fig.~\ref{dens300}, we see that due to the signal-recycling mirror the bright port fluctuations become less important. The reason is that we approximate at frequencies which are less than the (half-)bandwidth of the signal-recycling cavity $\gamma_{\rm sr}=200\,$Hz. For $\Omega\ll 2\pi\cdot \gamma_{\rm sr}$, the signal is weaker with signal-recycling cavity compared to a configuration without signal-recycling mirror. The bright port field behaves in the same way as the signal field whereas the fluctuations from the dark port are nearly unaffected by the signal-recycling mirror. Therefore the noise-to-signal ratio of the dark-port fluctuations is increased with respect to the noise-to-signal ratio of the bright-port field which is not changed by the signal-recycling mirror. Since the detuning of the signal-recycling cavity is very small, we find the following balance for the dual-recycled configuration GEO\,600
\beq
\sigma_{\rm GEO} \approx\frac{\tau_{\rm pr}}{\tau_{\rm sr}}\left(\frac{K_A}{2K_B}+1\right)=\frac{\tau_{\rm pr}}{\tau_{\rm sr}}\cdot\frac{m+m_M/5}{m_M/5}\approx 7.7
\label{estGEO}
\eeq
That value is in good agreement with the low frequency dark port and bright port vacuum noise in Fig.~\ref{dens300}. Inserting the respective values for the final set-up of GEO\,600, the balance is $\sigma_{\rm fGEO}\approx 15$ which also agrees with the spectral densities in Fig.~\ref{dens10k}. 

\section{THE NOISE SPECTRAL DENSITY FOR THE ADVANCED LIGO TOPOLOGY}
\begin{figure}[ht]
\begin{picture}(150,150)			
\put(0,40){\line(1,0){140}}
\put(40,0){\line(0,1){140}}
\linethickness{0.1cm}
\put(35,70){\line(1,0){10}}
\put(35,140){\line(1,0){10}}
\put(70,35){\line(0,1){10}}
\put(140,35){\line(0,1){10}}
\put(10,35){\line(0,1){10}}
\put(35,10){\line(1,0){10}}
\thicklines
\put(35,35){\line(1,1){10}}
\footnotesize
\put(3,50){PRM}
\put(63,50){ITM}
\put(133,50){ETM}
\put(48,7){SEM}
\put(48,67){ITM}
\put(48,137){ETM}
\end{picture}
\caption{The advanced LIGO configuration is a power-recycling, signal-extraction Michelson interferometer with arm cavities. The armlength is $4\,$km. Each arm cavity is formed by an input test mass and an end test mass. The signal-extraction cavity is formed by a mirror in the dark port and the interferometer which has the conventional LIGO topology.}
\label{LIGO}
\end{figure}
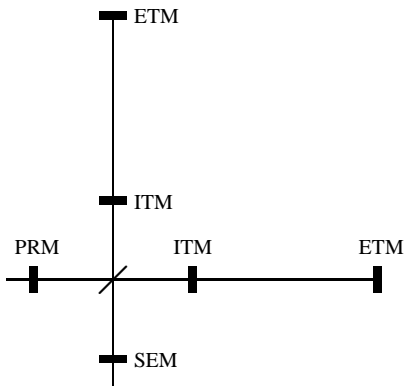

\begin{figure}[ht!]
\centerline{\includegraphics[width=8cm]{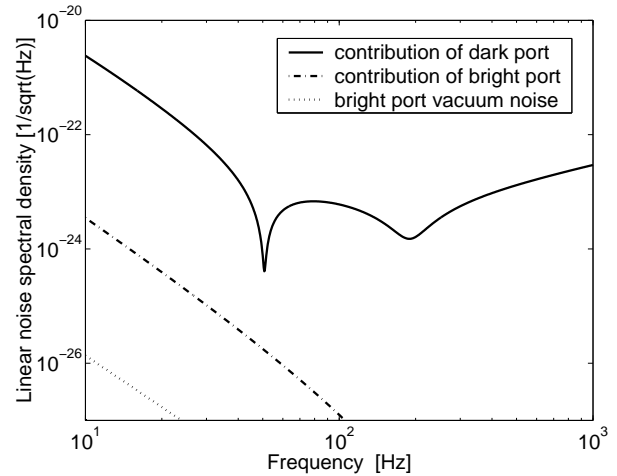}}
\caption{Single-sided spectral density for the LIGO topology with $P=10\,$kW at the beamsplitter. We chose the same spectral density of relative technical laser noise than for the GEO\,600 configurations. The bright-port fluctuations are negligible.}
\label{densLIGO}
\end{figure}

In this section, we apply the methods of the last section to the advanced LIGO configuration. According to the current plan \cite{refLIGO}, advanced LIGO will be a power-recycling, signal-extraction Michelson interferometer with arm cavities (see Fig.~\ref{LIGO}). 
The formulas which have to be applied for LIGO are identical to the formulas which were derived in the last section. The only difference lies in the definition of the arm coupling constant $K_A$ and the signal field in terms of the standard quantum limit which counts for the arm topology of LIGO (compare with Eq.~(\ref{couK}) and Eq.~(\ref{signal}))
\begin{eqnarray}
K_A & = & \frac{8\omega_0 P}{m_M L^2}\frac{1}{\Omega^2(\gamma^2+\Omega^2)} \\
h_{\rm SQL} & = & \frac{8\hbar}{m_M L^2 \Omega^2}
\end{eqnarray}
Also the additional phase shift gained by the modulation fields which are now reflected at the inner test mass of the arm cavity has to be replaced by an expression which depends on the half-bandwidth $\gamma=c\tau_{\rm itm}^2/4L$ of the arm cavity
\beq
\frac{\Omega L}{c} \longrightarrow \arctan\left(\frac{\Omega}{\gamma}\right)
\eeq
All parameter values which enter the preceding definitions are gathered from \cite{KLMTV01,BCh01,refLIGO}. They are listed in Tab.~\ref{parLIGO}. The mass of the beamsplitter is accurate up to some small percentage.
\begin{table}[ht!]
\begin{tabular}{l|cr}
\hline\hline
& Symbol & Value $\quad$\\
\hline
Light power at BS & $P$ & $10\,\rm kW\quad$\\
Transmissivity PRM $\quad$ & $\qquad\tau_{\rm pr}^2\qquad$ & $\quad 5\%\quad$ \\
Transmissivity SRM & $\tau_{\rm sr}^2$ & $19\%\quad$ \\
Transmissivity of ITM & $\tau_{\rm itm}^2$ & $3.3\%\quad$ \\
Beamsplitter mass & $m$ & $13\,\rm kg \quad$ \\
Mirror mass & $m_M$ & $40\,\rm kg \quad$ \\
Arm length & $L$ & $4000\,\rm m \quad$ \\
Frequency of laser & $\omega_0$ & $1.77\cdot 10^{15}\,{\rm rad/s}\quad$ \\
Detuning of SE cavity & $\phi$ & $\pi/2-0.47\,\rm rad\quad$ \\
\hline\hline
\end{tabular}
\caption{Parameters of the advanced LIGO configuration. Except for the beamsplitter mass, the values of the parameters are chosen according to \cite{KLMTV01,BCh01,refLIGO}. The transmissivity of the power-recycling mirror corresponds to a power amplification factor of 80 and so the input light power has to be $125\,$W.}
\label{parLIGO}
\end{table}
The noise spectral density of the output field shown in Fig.~\ref{densLIGO} lies well above the spectral density of the bright port vacuum and also above the technical noise from the bright port. Again the latter one is characterized by the same relative technical noise than in the two cases discussed for GEO\,600. The fact that the bright-port dark-port coupling is insignificant for LIGO was anticipated and can be further quantified by performing a comparison of the beamsplitter and arm coupling constants.
Due to the increased power in the arm cavities, LIGO's arm coupling constant $K_A$ is much bigger than the beamsplitter coupling $K_B$. At low frequencies, the coupling constant $K_A$ can be approximated by 
\beq
K_A=\frac{8\omega_0 P}{mc^2\Omega^2}\left(\frac{4}{\tau^2_{\rm itm}}\right)^2
\eeq
Evaluating the ratio of the two matrix components $K_1$, $K_2$ of Eq.~(\ref{GEOio}) for a modulation frequency $\Omega=2\pi\cdot10\,$Hz which is far below the half-bandwidth of the arm cavities, we obtain
\beq
\sigma_{\rm LIGO}\approx\frac{2\Omega}{\tau_{\rm pr}\gamma}\left(\frac{K_A}{2K_B}+1\right)\approx\frac{8\Omega}{\tau_{\rm pr}\gamma}\frac{m}{m_M}\left(\frac{4}{\tau^2_{\rm itm}}\right)^2\approx 8\cdot 10^4.
\eeq
Comparing with Eq.~(\ref{estGEO}), the result shows that the low frequency balance is five orders of magnitude bigger for LIGO than for GEO\,600 corresponding to a weaker bright-port contribution to the output field. Even if the technical fluctuations of the input light are two or three orders of magnitude stronger than pure vacuum fluctuations, there will be no noticable contribution to the spectral density of the output field.

\section{CONCLUSION}
We showed that the bright-port dark-port coupling gives rise to a significant contribution of technical fluctuations to the noise spectral density at low frequencies for a Michelson topology without arm cavities. In terms of the low frequency balance, we concluded that the LIGO topology exhibits a comparatively weak bright-port dark-port coupling relative to the contribution of the dark port noise. That is true even for a high level of technical laser noise. There are a couple of strategies to reduce these fluctuations at the dark port of the GEO\,600 topology. One option is to decrease the transmissivity of the power-recycling mirror. The proposition seems to be in contradiction to Eq.~(\ref{estGEO}) which states that the relative bright-port fluctuations increase with decreasing $\tau_{\rm pr}$. The reason why it works is, that the factor $\tau_{\rm pr}/2$ in front of the brackets has to be replaced by $\Omega L/(\tau_{\rm pr}c)$ if the following condition holds: $\tau_{\rm pr}^2\ll \Omega L/c$. The required amplitude transmissivity had to be around 10ppm which lies beyond any practical feasibility and which is not desired for other reasons. The most obvious option is to increase the mass of the beamsplitter without increasing the masses of the endmirrors. One can see from Eq.~(\ref{estGEO}) that the beamsplitter mass has to be increased by two to three orders of magnitude depending on the technical noise which seems to be unfeasible again. The results of this paper do not have any practical relevance for currently operated interferometers whose sensitivity at low frequencies is limited not by the technical noise from the bright port, but either by seismic noise or thermal noise. The lasers which furnish the light for interferometers of the next generation are supposed to have a considerably lower amount of relative technical noise. Therefore, increasing the mass of the beamsplitter by a modest factor might already be sufficient in order to make the bright port fluctuations negligible for the GEO\,600 topology. Also strategies which yield an increase of the sensitivity at high frequencies \cite{Sch04} are not influenced by our results.

There is another generic mechanism existing by which a bright-port dark-port coupling is built up. If the transfer function of the two arms are not equal, then the input-output relation contains the following contribution from the bright port
\beq
{\bf\bar o}={\bf IO}({\bf\bar i},{\bf\bar n},{\bf\bar e})+\frac{\rho\tau (N-E)}{1-(\rho^2\, N+\tau^2\,E)\cdot W}\cdot {\bf\bar w}
\label{alter}
\eeq
where $\rho$ and $\tau$ denote the amplitude reflectivity and transmissivity of the beamsplitter. There are different reasons why the two arms are not described by the same transfer function. One reason could be that one arm is detuned intentionally in order to transmit some carrier light towards the detector where it serves as local oscillator for a homodyne measurement of the signal. For a small detuning of just one arm, the carrier light is transmitted into the phase quadrature of ${\bf\bar o}$ governed by a transfer function which is proportional to the detuning. If the carrier light is to be transmitted into the amplitude quadrature of ${\bf\bar o}$, then one can choose the detuning of one arm to be $\pi/2$ and the detuning of the other arm to be slightly less than $\pi/2$. In that case, the west port has to be detuned accordingly to maintain the power-recycling condition (which cannot hold exactly here, since the two arms are detuned relative to each other). An unintentional reason for different transfer functions of the two arms could be that the transmissivity and the reflectivity of the beamsplitter are not the same. Then, radiation-pressure fluctuations in the two arms would be different by virtue of the different powers of the two respective carrier fields. Carrier light is then transmitted into the phase quadrature of ${\bf\bar o}$ and the corresponding transfer function is proportional to the difference of the power reflectivity and transmissivity of the beamsplitter. There would also be a small loss of the optical signal due to a partial transmission into the bright port which is proportional to the same difference of power reflectivity and transmissivity. 

\section{ACKNOWLEDGMENTS}
We thank Y.~Chen whose invaluable suggestions and scrutiny contributed to clarify the paper. We also acknowledge M.~Heurs, H.~Grote and B.~Willke for
providing us with all the required details of the GEO\,600 interferometer. We thank W.~Winkler for his encouragement towards a quantitative and quantum mechanical treatment of the finite mass beamsplitter.


\begin{thebibliography}{12}
\bibitem{geo02} B.~Willke et al., Class. Quantum Grav. {\bf 19}, 1377 (2002)
\bibitem{LIGO} D.~Sigg, Class.~Quantum Grav.~ {\bf 21}, S409 (2004)
\bibitem{TAMA} M.~Ando et al., Phys.~Rev.~Lett. {\bf 86}, 3950 (2001)
\bibitem{VIRGO} B.~Caron et al., Class.~Quantum Grav. {\bf 14} 1461 (1997)
\bibitem{Cav80} C.~M.~Caves, Phys.~Rev.~Lett. {\bf 45}, 75 (1980)
\bibitem{KLMTV01} H.~J.~Kimble, Y.~Levin, A.~B.~Matsko, K.~S.~Thorne and S.~P.~Vyatchanin, Phys. Rev. D {\bf 65}, 022002 (2001)
\bibitem{BCh01} A.~Buonanno and Y.~Chen, Phys.~Rev.~D {\bf 64}, 042006 (2001)
\bibitem{Cav81} C.~M.~Caves, Phys.~Rev.~D {\bf 23}, 1693 (1981)
\bibitem{JH03} J.~Harms, Y.~Chen, S.~Chelkowski, A.~Franzen, H.~Vahlbruch, K.~Danzmann and R.~Schnabel, Phys. Rev. D {\bf 68}, 042001 (2003)
\bibitem{Sch04} R.~Schnabel, J.~Harms, K.~A.~Strain and K.~Danzmann, Class.~Quantum Grav.~{\bf 21}, S1045 (2004)
\bibitem{CSc85} C.~M.~Caves and B.~L.~Schumaker, Phys.~Rev.~A, {\bf 31}, 3068 (1985) 
\bibitem{PC02} P.~Purdue and Y.~Chen, Phys.~Rev.~D {\bf 66}, 122004 (2002)
\bibitem{Ch03} Y.~Chen, Phys.~Rev.~D {\bf 67}, 122004 (2003) 
\bibitem{BCh02} A.~Buonanno and Y.~Chen, Phys.~Rev.~D {\bf 65}, 042001 (2002)
\bibitem{HarmsDipl} J.\,Harms, \emph{Quantum Noise in the Laser-Interferometer Gravitational-Wave Detector GEO\,600}, Diploma thesis, Universit\"at Hannover, 2002, available at http://www.geo600.uni-hannover.de/personal/harms.html
\bibitem{BChMa} A.~Buonanno, Y.~Chen and N.~Mavalvala, Phys.~Rev.~D {\bf 67}, 122005 (2003)
B.~L.~Schumaker and C.~M.~Caves, Phys.~Rev.~A {\bf 31}, 3093 (1985)
\bibitem{refLIGO} A.~Weinstein, Class.~Quantum Grav.~{\bf 19}, 1575 (2002)
\bibitem{S3run} S3 is the name for the third sience run of the LIGO detectors which was joined by GEO\,600
\end{thebibliography}
\end{document}